\documentclass[12pt,epsf]{article}
\usepackage{amssymb,amsmath}
\usepackage{graphicx}

\newcommand{\be}{\begin{equation}}
\newcommand{\ee}{\end{equation}}
\newcommand{\bea}{\begin{eqnarray}}
\newcommand{\eea}{\end{eqnarray}}
\newcommand{\bear}{\begin{eqnarray}}
\newcommand{\eear}{\end{eqnarray}}
\newcommand{\beas}{\begin{eqnarray*}}

\newcommand{\eeas}{\end{eqnarray*}}
\newcommand{\ba}{\begin{array}}
\newcommand{\ea}{\end{array}}

\newcommand{\tr}{\operatorname{tr}}
\newcommand{\pd}[2][1]{\ifnum#1=1 \frac{\partial}{\partial {#2}} \else
  \frac{\partial^#1}{\partial {#2}^{#1}}\fi}
\newcommand{\dpd}[2][1]{\ifnum#1=1 \dfrac{\partial}{\partial {#2}} \else
  \frac{\partial^#1}{\partial {#2}^{#1}}\fi}
\newcommand{\td}[2][1]{\ifnum#1=1 \frac{d}{d{#2}} \else
  \frac{d^#1}{d{#2}^{#1}}\fi}

\newcommand{\nbox}{{\,\lower0.9pt\vbox{\hrule \hbox{\vrule height 0.2 cm \hskip 0.19 cm \vrule height 0.2 cm}\hrule}\,}}

\def\href#1#2{#2}

\begin{document}
\begin{titlepage}
\hfill
\vbox{
    \halign{#\hfil         \cr
           } 
      }  
\vspace*{20mm}
\begin{center}
{\Large \bf Gravitational Dynamics \\ From Entanglement ``Thermodynamics''}

\vspace*{15mm}
\vspace*{1mm}
Nima Lashkari, Michael B. McDermott, Mark Van Raamsdonk
\vspace*{1cm}

{Department of Physics and Astronomy,
University of British Columbia\\
6224 Agricultural Road,
Vancouver, B.C., V6T 1W9, Canada}

\vspace*{1cm}
\end{center}
\begin{abstract}

In a general conformal field theory, perturbations to the vacuum state obey the relation $ \delta S =  \delta E$, where $ \delta S$ is the change in entanglement entropy of an arbitrary ball-shaped region, and $ \delta E$ is the change in ``hyperbolic'' energy of this region. In this note, we show that for holographic conformal field theories, this relation, together with the holographic connection between entanglement entropies and areas of extremal surfaces and the standard connection between the field theory stress tensor and the boundary behavior of the metric, implies that geometry dual to the perturbed state satisfies Einstein's equations expanded to linear order about pure AdS.

\end{abstract}

\end{titlepage}

\vskip 1cm

\section{Introduction}

Since the first connections between gravity and thermodynamics were realized in the study of black hole physics \cite{Bekenstein:1973ur, Bardeen:1973gs, Hawking:1974sw}, various attempts have been made to derive Einstein's equations from the thermodynamics of some underlying degrees of freedom, starting with Jacobson's intriguing paper \cite{Jacobson:1995ab} (see also \cite{Padmanabhan:2009vy, Verlinde:2010hp}). With the AdS/CFT correspondence \cite{malda,agmoo}, the underlying degrees of freedom for certain theories of gravity with AdS asymptotics have been explicitly identified as the degrees of freedom of a conformal field theory. It is thus interesting to ask whether the Einstein's equations in the gravitational theory can be derived from some thermodynamic relations for the CFT degrees of freedom.

In this note, following \cite{Ryu:2006bv,Hubeny:2007xt,Casini:2011kv,Nozaki:2013vta,Blanco:2013joa} we demonstrate that at least to linear order in perturbations around pure AdS, Einstein's equations do follow from a relation $dE = dS$ closely related to the First Law of Thermodynamics, but where the entropy $S$ is the {\it entanglement entropy} of a spatial region in the field theory, and $E$ is a certain energy associated with this region. A key point is that $dS$ and $dE$ can be defined and the relation $dS=dE$ shown to hold for {\it arbitrary} perturbations around the vacuum state; thus, the relation is more general than the ordinary first law which applies only in situations of thermodynamic equilibrium.

The specific relation we employ, which we write as
\be
\label{main}
\delta S_A = \delta E^{hyp}_A
\ee
was derived recently by Blanco, Casini, Hung, and Myers in \cite{Blanco:2013joa}. Here $A$ represents a ball-shaped spatial region, $\delta S_A$ represents the change in entanglement entropy of the region $A$ relative to the vacuum state, and $\delta E^{hyp}_A$ represents the ``hyperbolic'' energy of the perturbed state in the region $A$, the expectation value of an operator which maps to the Hamiltonian of the CFT on hyperbolic space times time under a conformal transformation that takes the domain of dependence of the region $A$ to $H^d \times {\rm time}$. We review the derivation of this relation in section 2 below.

For holographic conformal field theories, each side of (\ref{main}) has an interpretation in the dual gravity theory. Assuming that the perturbed state $| \Psi \rangle$ corresponds to some weakly-curved classical spacetime, the entanglement entropy $S_A$ may be calculated via the Ryu-Takayanagi proposal \cite{Ryu:2006bv} and its covariant generalization \cite{Hubeny:2007xt} as the area of an extremal surface in the bulk, as we review in section 3.1. In section 3.2, we recall that the energy $\delta E_A$ can be calculated from the asymptotic behavior of the metric. Thus, the field theory relation $\delta S_A = \delta E^{hyp}_A$ translates to a constraint on the dual geometry.

In section 4, we show that this constraint is precisely that the bulk metric corresponding to $|\Psi \rangle$ must satisfy Einstein's equations to linear order in the perturbation around pure AdS (the geometry corresponding to the CFT vacuum state). That solutions of Einstein's equations satisfy $\delta S_A = \delta E^{hyp}_A$ has already been shown in \cite{Blanco:2013joa} (see also the related earlier work \cite{Nozaki:2013vta, Allahbakhshi:2013rda, Wong:2013gua}). For completeness, we provide an alternate demonstration of this in section 4.2. In section 4.3, we go the other direction, showing that any perturbation to pure AdS satisfying $\delta S_A = \delta E^{hyp}_A$ must satisfy Einstein's equations. This requires more than simply reversing the arguments of section 4.2 (or of \cite{Blanco:2013joa}). In particular, demanding that $\delta S_A = \delta E^{hyp}_A$ for all ball-shaped spatial regions $A$ in a particular Lorentz frame only places mild constraints on the metric, determining the combination $H_{xx} + H_{yy}$ in terms of the other
components. It is only when we demand that $\delta S_A = \delta E^{hyp}_A$ in an arbitrary Lorentz frame (i.e. for ball-shaped regions on arbitrary spatial slices) that the full set of linearized Einstein's equations is implied.

In appendix A, we give an alternative proof that Einstein's equations imply $\delta S_A = \delta E^{hyp}_A$ that is perhaps more straightforward, but assumes that the metric is analytic.

We conclude in section 5 with a discussion.

\section{Entropy-energy relation}\label{introsection}

In this section, we review the relation $\delta S_A = \delta E^{hyp}_A$, derived by Blanco, Casini, Hung, and Myers in \cite{Blanco:2013joa} as a special case of an inequality that follows from the positivity of relative entropy.

\subsubsection*{General expression for variation of the entanglement entropy}

Consider a CFT on $R^{d,1}$ in some state $|\Psi \rangle$. Choosing a spatial region $A$, define $\rho_A$ to be the reduced density matrix associated with this region for the state $|\Psi \rangle$,
\[
\rho_A = \tr_{\bar{A}}|\Psi\rangle \langle \Psi| \; .
\]
From this, we can define the modular Hamiltonian $H_A$ by
\[
\rho_A = e^{-H_A} \; .
\]
For general states, this modular Hamiltonian is not related to the usual Hamiltonian, and cannot be written as the integral of a local density. We now consider an arbitrary variation of the state $|\Psi \rangle$. The change in entanglement entropy $S_A$ for the region $A$ is given by
\beas
\delta S_A &=& \delta ( - \tr(\rho_A \log \rho_A)) \cr
&=& -\tr(\delta \rho_A \log \rho_A) \cr
&=& \tr(\delta \rho_A H_A) \cr
&=& \delta \langle H_A \rangle
\eeas
where we have used the fact that $\tr(\delta \rho_A) = 0$, a consequence of assuming that the density matrix has a fixed normalization. In the last line, $H_A$ is the original modular Hamiltonian associated with the density matrix $\rho_A$ for the original state. Thus, we have the general relation
\be
\label{general}
\delta S_A = \delta \langle H_A \rangle \; ,
\ee
valid in any spatial region $A$ for arbitrary perturbations of an arbitrary state.

\subsubsection*{``Thermodynamic'' relation for perturbations around the vacuum state}

We now specialize to the case where $|\Psi \rangle$ is the vacuum state, and the region $A$ is a ball of radius $R$. In this case, the domain of dependence of the ball-shaped region\footnote{The domain of dependence of $A$ is the set of points $p$ for which all inextensible causal curves passing through $p$ also pass through $A$.} can be mapped by a conformal transformation to hyperbolic space times time. As shown in \cite{Casini:2011kv}, such a transformation maps the vacuum density matrix for the region $A$ to the thermal density matrix $e^{-\beta H_{hyp}}$ for the hyperbolic space theory, where the temperature is related to the hyperbolic space curvature radius $R_H$ by $\beta = 2 \pi R_{hyp}$. In this case $H_{hyp}$ is the integral of the local operator $T^{00}_{hyp}$ over hyperbolic space. Mapping back to the ball-shaped region of Minkowski space, it follows \cite{Casini:2011kv} that the modular Hamiltonian can be written as
\[
H^{vac}_A = 2 \pi \int_A d^d x {R^2 - r^2 \over 2 R} T^{00}
\]
where $T^{00}$ is the energy density operator for the CFT and $r$ is a radial coordinate centered at the center of the ball.

In this case, we have
\be
\label{mod}
\delta \langle H_A \rangle = 2 \pi \int_A d^d x {R^2 - r^2 \over 2 R} \delta T^{00} \equiv \delta E^{hyp}_A \; ,
\ee
i.e. the variation in the expectation value of the vacuum modular Hamiltonian $H^{vac}_A$ under a small perturbation away from the vacuum state is equal to the change in the ``hyperbolic'' energy of the region. Thus, the general relation (\ref{general}) gives
\be
\label{predict}
\delta S_A = \delta E^{hyp}_A \; ,
\ee
reminiscent of the First Law of Thermodynamics. We emphasize however that the entanglement entropy $S_A$ can be defined for any state, in contrast to the usual thermodynamic entropy which applies to equilibrium states. Thus, (\ref{predict}) represents a much more general result.

\section{Gravitational implications of $dS=dE$ in holographic theories}

Let us now consider the case of a holographic conformal field theory on Minkowski space, whose states correspond to asymptotically AdS spacetimes in some quantum theory of gravity. In this case, each side of the relation $\delta S_A = \delta E^{hyp}_A$ has a straightforward gravitational interpretation. As we review below, the left side may be calculated using the Ryu-Takayanagi proposal \cite{Ryu:2006bv,Hubeny:2007xt}, while the right side can be calculated from the asymptotic form of the metric. The equality of these quantities represents a constraint on the gravitational dynamics implied by the dual field theory. In the next section, we show that this constraint is precisely equivalent to Einstein's equations linearized about AdS.

\subsection{Gravitational calculation of $dS$}

According to the Ryu-Takayanagi proposal \cite{Ryu:2006bv} and its covariant generalization \cite{Hubeny:2007xt}, the entanglement entropy $S_A$ for a state with a geometrical gravity dual is proportional to the area of the extremal co-dimension two surface $\tilde{A}$ in the bulk whose boundary coincides with the boundary of the region $A$ on the AdS boundary,
\[
S_A = {\rm Area(\tilde{A}) \over 4 G_N} \; .
\]
The surface $\tilde{A}$ is an extremum of the area functional
\[
A(G,X_{ext}) = \int d^d \sigma \sqrt{g}
\]
where
\[
g = \det(g_{ab}) = \det(G_{\mu \nu}) {d X^\mu \over d \sigma^a} {d X^\nu \over d \sigma^b} \; .
\]
Starting from pure AdS, with metric\footnote{Throughout this paper, we set the AdS radius to one.}
\be
\label{AdS}
ds^2 = G^{0}_{\mu \nu} dx^\mu dx^\nu  = {1 \over z^2}(-dt^2 + dz^2 + d \vec{x}^2)
\ee
the extremal surface ending on the spatial boundary sphere of radius $R$ is described by the spacetime surface
\be
\label{extremal}
\vec{x}^2 + z^2 = R^2 \; .
\ee

We now consider a small variation
\be
\label{metricpert}
G_{\mu \nu} = G^{0}_{\mu \nu} + \delta G_{\mu \nu} \; .
\ee
In this case, the extremal surface changes, and the new area is
\[
A(G_0 + \delta G,X^0_{ext} + \delta X)
\]
where the variation $\delta X$ will be of order $\delta G$. Since the original surface was extremal, we have
\[
A(G_0,X^0_{ext} + \delta X) = A(G_0,X^0_{ext}) + {\cal O}(\delta X^2) \; .
\]
Thus, the variation of the surface gives rise to changes in the area that start at order $\delta G^2$. To find the order $\delta G$ variation of the area, we need only evaluate
\[
A(G_0 + \delta G,X^0_{ext}) - A(G,X^0_{ext})
\]
expanded to linear order in $\delta G$. We find that
\be
\label{dSint0}
\delta A = \int d^d \sigma {1 \over 2} \sqrt{g_0} g_0^{ab} \delta g_{ab} \; ,
\ee
where we have used lower-case letters to represent pullbacks to the extremal surface. Thus, for field theory state $|\Psi \rangle$ close to the vacuum state with dual geometry described by (\ref{metricpert}), the change in the entanglement entropy for region $A$ relative to the vacuum state is given by an integral of the metric perturbation over the original extremal surface $\tilde{A}$. Using the explicit metric (\ref{AdS}) and parameterizing the extremal surface (\ref{extremal}) by the boundary coordinates $x^i$, we have finally that
\be
\label{dSint}
\delta S = {R \over 8 G_N}\int d^d x  (\delta_{ij} - {1 \over R^2} x_i x_j) H_{ij} \; .
\ee

\subsection{Gravitational calculation of $dE$}

General asymptotically AdS spacetimes with a Minkowski space boundary geometry may by described using Fefferman-Graham coordinates by a metric
\be
\label{FG}
ds^2 = {1 \over z^2}(dz^2 + dx_\mu dx^\mu + z^d  H_{\mu \nu}(x,z) dx^\mu dx^\nu) \; .
\ee
where pure AdS, dual to the CFT vacuum, corresponds to $H_{\mu \nu} = 0$. With this parametrization, the expectation value $t_{\mu \nu}$ of the field theory stress-energy tensor is simply related to the asymptotic metric by \cite{Myers:1999psa,de Haro:2000xn}
\[
 t_{\mu \nu} (x)  = {d \over 16 \pi G_N} H_{\mu \nu}(z=0,x) \; .
\]
Thus, we may write the change in the hyperbolic energy (\ref{mod}) relative to the vacuum state as
\be
\label{dEint}
\delta E_A^{hyp} =  {d \over 16 G_N} \int_A d^d x {R^2 - r^2 \over R} \delta H_{00} (0,x)  \; .
\ee
This is an integral of the boundary value of $H$ over the region $A$.

\section{Derivation of linearized Einstein's equations from $dE=dS$}

We are now ready to demonstrate that using the holographic dictionary reviewed in the previous section, the CFT relation $\delta S_A = \delta E^{hyp}_A$ is equivalent to the constraint that metric corresponding to the perturbed CFT satisfies Einstein's equations to linear order. For clarity, we focus on the case of 2+1 dimensional conformal field theories, corresponding to gravitational theories with four non-compact dimensions. However, the result can also be proven for general higher-dimensional theories.

Using the results (\ref{dSint}) and (\ref{dEint}), the CFT relation $\delta S_A = \delta E^{hyp}_A$ implies that a disk of any radius $R$ centered at any point $(x_0,y_0)$ on the boundary, the integral
\bea
\label{LS}
\delta \hat{S} &=& \int_{D_R} dx dy  \left\{H_{xx}(\sqrt{R^2 - x^2 - y^2},t,x+x_0,y+y_0)(R^2 - x^2) \right. \cr
 &&\qquad \qquad \qquad  + H_{yy}(\sqrt{R^2 - x^2 - y^2},t,x+x_0,y+y_0)(R^2 - y^2)\cr
 &&\qquad \qquad \qquad  \left. - 2 H_{xy}(\sqrt{R^2 - x^2 - y^2},t,x+x_0,y+y_0)xy \right\}
\eea
over the bulk extremal surface must equal the integral
\be
\label{RS}
\delta \hat{E}= {3 \over 2} \int_{D_R} dx dy (R^2 - x^2 - y^2) H_{tt}(0,t,x+x_0,y+y_0) \;
\ee
over the $z=0$ surface, where we have absorbed a factor of $1/8 G_N R$ to define $\delta \hat{S}(R,x_0,y_0)$ and $\delta \hat{E}(R,x_0,y_0)$ (we drop the hats from now on). We will now show that this equality is true for all disks {\it in all Lorentz frames} if and only if the bulk metric satisfies Einstein's equations to linear order in $H$. As shown in \cite{Blanco:2013joa}, these are equivalent to the set of equations
\be
\label{Einstein}
H_\alpha {}^\alpha = 0 \qquad \qquad \partial_\mu H^{\mu \nu} = 0 \qquad \qquad {1 \over z^4} \partial_z \left\{z^4 \partial_z H_{\mu \nu} \right\} + \partial^2 H_{\mu \nu}  = 0 \;
\ee
that arise by plugging the Fefferman-Graham form of the metric (\ref{FG}) into the $zz$, $z \mu$, and $\mu \nu$ components of Einstein's equations
\[
W_{\mu \nu} = R_{\mu \nu} - {1 \over 2} g_{\mu \nu} R - 3 g_{\mu \nu} = 0 \; ,
\]
respectively and using the fact that $H$ is regular at $z=0$. In (\ref{Einstein}), the last equation is equivalent to saying that each component of $z^3 H$ must satisfy the Laplace equation on the AdS background.

\subsection{Proof that $\delta S = \delta E$ for solutions of Einstein's equations}

We begin by showing that solutions of the linearized Einstein's equations obey the equality $\delta S = \delta E$. This has already been checked in section 3.1 of \cite{Blanco:2013joa} by demonstrating the result for a complete basis of solutions to the equations (\ref{Einstein}). In this section, we offer an alternative proof that does not require using an explicit basis of solutions. A third proof that is perhaps more straightforward but assumes a series expansion of $H$ is given in appendix A.

Using the equations (\ref{Einstein}), we have:
\be
\ba{rl}
& \partial_t^2 H_{tt} = \partial_t^2 (H_{xx} + H_{yy}) \qquad \qquad  \cr
\Rightarrow & \partial_t (\partial_x H_{xt} + \partial_y H_{yt})  = \partial_t^2 (H_{xx} + H_{yy})  \cr
\Rightarrow & \partial_x^2 H_{xx} + \partial_y^2 H_{yy} + 2 \partial_x \partial_y H_{xy}  = \partial_t^2 (H_{xx} + H_{yy})   \cr
\Rightarrow & \partial_x^2 H_{xx} + \partial_y^2 H_{yy} + 2 \partial_x \partial_y H_{xy}  = \partial_t^2 (H_{xx} + H_{yy})   \cr
\Rightarrow & \partial_x^2 H_{xx} + \partial_y^2 H_{yy} + 2 \partial_x \partial_y H_{xy}  = (\partial_x^2 + \partial_y^2) (H_{xx} + H_{yy}) + {1 \over z^4} \partial_z(z^4 \partial_z (H_{xx} + H_{yy}))   \cr
\Rightarrow & 2 \partial_x \partial_y H_{xy}  = \partial_y^2 H_{xx} + \partial_x^2 H_{yy} + {1 \over z^4} \partial_z(z^4 \partial_z (H_{xx} + H_{yy}))
\ea
\label{Hxysolve}
\ee
We would like to use the last equation to eliminate $H_{xy}$ from (\ref{LS}). However, we have $H_{xy}$ rather than $\partial_x \partial_y H_{xy}$ in (\ref{LS}). To make progress, we begin by differentiating $\delta S$ by $x_0$ and $y_0$ (the coordinates of the center of the boundary disk). This gives
\bea
\label{LSxy}
\partial_{x_0} \partial_{y_0} \delta S &=&  \int_{D_R} dx dy  \left\{\partial_{x} \partial_{y} H_{xx}(\sqrt{R^2 - x^2 - y^2},t,x+x_0,y+y_0)(R^2 - x^2) \right. \cr
 &&\qquad \qquad \qquad  + \partial_{x} \partial_{y} H_{yy}(\sqrt{R^2 - x^2 - y^2},t,x+x_0,y+y_0)(R^2 - y^2)\cr
 &&\qquad \qquad \qquad  \left. - 2 \partial_{x} \partial_{y} H_{xy}(\sqrt{R^2 - x^2 - y^2},t,x+x_0,y+y_0) x y  \right\}
\eea
Now, using (\ref{Hxysolve}), we have
\bea
\label{LSxy}
\partial_{x_0} \partial_{y_0} \delta S &=& \int_{D_R} dx dy  \left\{\partial_{x} \partial_{y} H_{xx}(R^2 - x^2 )  + \partial_{x} \partial_{y} H_{yy}(R^2 - y^2) \right.\cr
 && \qquad \qquad \qquad \left. - xy \left( \partial_y^2 H_{xx} + \partial_x^2 H_{yy} + {1 \over z^4} \partial_z(z^4 \partial_z (H_{xx} + H_{yy})) \right) \right\}
\eea
It is straightforward to check that this expression is equal to the integral over the extremal surface of an exact form $dA$, where $A$ is defined for all $(x,y,z,t)$ as
\bea
\label{defA}
A &=& \left( -xz \partial_z H_{xx} - 3 x H_{xx} + z^2 \partial_x H_{yy} \right) dx \cr
&&+ \left(z^2 \partial_y H_{xx} - yz \partial_z H_{yy} - 3 y  H_{yy} \right) dy \cr
&&+ \left( - y z \partial_y H_{xx} - xz \partial_x H_{yy} \right) dz \; .
\eea
By Stokes theorem, this equals the integral of $A$ over the boundary of the extremal surface, so we have
\beas
\partial_{x_0} \partial_{y_0} \delta S &=&  \int_{\partial D_R} A \cr
&=& -3 \int_{\partial D_R}( x H_{xx} dx + y H_{yy} dy ) \cr
&=& 3 \int d \theta (H_{xx} + H_{yy}) \cos(\theta) \sin(\theta)
\eeas
In the second step, we have used the fact that all other terms in $A$ vanish for $z=0$.

Similarly, we find that $\partial_{x_0} \partial_{y_0} \delta E$ may be written as
\beas
\partial_{x_0} \partial_{y_0} \delta E &=& {3 \over 2} \int_{D_R} dx dy  \partial_{x_0} \partial_{y_0} H_{tt}(0,t,x+x_0,y+y_0) (R^2 - x^2  - y^2) \cr
&=& {3 \over 2} \int_{D_R} dx dy  \partial_{x} \partial_{y} (H_{xx}(0,t,x+x_0,y+y_0) + H_{yy}(0,t,x+x_0,y+y_0)) (R^2 - x^2  - y^2) \cr
&=& {3 \over 2} \int_{D_R} d\hat{A} \; ,
\eeas
where we can choose
\[
\hat{A} = \left(-2 x  H_{xx} + (R^2 - x^2  - y^2) \partial_x H_{yy} \right)dx +  \left(-2 y H_{yy} + (R^2 - x^2  - y^2) \partial_y H_{xx} \right) dy \; .
\]
Again, using Stokes theorem, this reduces to the integral of $(3/2) \hat{A}$ over the boundary, so
\beas
\partial_{x_0} \partial_{y_0} \delta E &=& {3 \over 2} \int_{\partial D_R} \hat{A} \cr
&=& - 3 \int_{\partial D_R} ( x H_{xx} dx + y H_{yy} dy ) \cr
&=& \partial_{x_0} \partial_{y_0} \delta S
\eeas
We conclude that for any $H$ satisfying Einstein's equations,
\[
\delta S (x_0, y_0, R ; H) - \delta E (x_0, y_0, R ; H) = C_x(x_0, R ; H) + C_y(y_0, R ; H) \; ,
\]
where $C_x$ and $C_y$ are some functionals linear in $H$ that do not depend on $y_0$ or $x_0$ respectively. Now, consider the class of functions $H$ that vanish for sufficiently large $x_0^2 + y_0^2$ at the time $t=0$ where we evaluate $\delta S$ and $\delta E$. In this case, fixing any $x_0$ and taking $y_0 \to \infty$ or fixing any $y_0$ and taking $x_0 \to \infty$, the left side must vanish. For this to be true on the right side, both $C_x$ and $C_y$ must be constant (as functions of $x_0$ and $y_0$), with $C_x + C_y = 0$. Thus, the right side vanishes for any $H$ that vanishes as $x_0^2 + y_0^2 \to \infty$. But more general $H$ can be written as linear combinations of such functions, and since the right side is a linear functional in $H$, it must vanish for all $H$. This completes the argument that $\delta S_A = \delta E^{hyp}_A$ for solutions of Einstein's equations.

\subsection{Proof that $\delta S = \delta E$ implies the linearized Einstein's equations}

In this section, we go the other direction to show that the relation $\delta S = \delta E$ implies that the metric satisfies Einstein's equations to linear order, i.e. that the equivalence of (\ref{LS}) and (\ref{RS}) implies the relations (\ref{Einstein}).

Given the boundary stress tensor $t_{\mu \nu}$, let $H^{EE}_{\mu \nu}$ be the corresponding metric perturbation that follows from Einstein's equations, i.e. the solution of (\ref{Einstein}) satisfying $H^{EE}_{\mu \nu}(0,t,x,y) = ( 16 \pi G_N /3 ) t_{\mu \nu}$. We will show that there is no other $H$ with these boundary conditions for which $\delta S = \delta E$ in all frames of reference.

Suppose there were another $H$ for which $\delta S = \delta E$ for all disk shaped regions in all Lorentz frames. Then the difference $\Delta = H - H^{EE}$ must satisfy
\be
\label{bound}
\Delta_{\mu \nu}(z=0,t,x,y) = 0 \; ,
\ee
and
\bea
\label{dspert}
0 &=& \int_{D_R} dx dy  \left\{\Delta_{xx}(\sqrt{R^2 - x^2 - y^2},x+x_0,y+y_0)(1 - {x^2 \over R^2}) \right. \cr
 &&\qquad \qquad \qquad  + \Delta_{yy}(\sqrt{R^2 - x^2 - y^2},x+x_0,y+y_0)(1 - {y^2 \over R^2})\cr
 &&\qquad \qquad \qquad  \left. - 2 \Delta_{xy}(\sqrt{R^2 - x^2 - y^2},x+x_0,y+y_0){xy \over R^2} \right\} \;
\eea
for arbitrary $R$, $x_0$, and $y_0$, and in an arbitrary Lorentz frame.

Let us first see the consequences of demanding this result in a fixed frame. To begin, we note that (\ref{dspert}) may be expanded in powers of $R$ using the basic integral
\[
\int_{D_R} dx dy (R^2 - x^2 - y^2)^{n \over 2} x^{2 m_x} y^{2 m_y}   =   R^{n + 2 m_x + 2 m_y + 2} I_{n,m_x,m_y}  \; ,
\]
where
\be
\label{defI}
I_{n,m_x,m_y} = {\Gamma(m_x + {1 \over 2}) \Gamma(m_y + {1 \over 2})  \Gamma({n \over 2} +1) \over \Gamma({n \over 2} + m_x + m_y + 2)} \; .
\ee
Defining
\be
\label{Delexp}
\Delta_{\mu \nu}(z,x,y) = \sum_{n=0}^\infty z^n \Delta_{\mu \nu}^{(n)}(x,y)
\ee
we find that (\ref{dspert}) becomes\footnote{Here, we are assuming that the function $\Delta$ is analytic. It would be useful to find a derivation of our result that holds more generally.}
\be
\label{ds2}
\ba{rl} 0 =  \sum R^{n + 2 m_x + 2 m_y + 2} \Big\{ &    {1  \over (2m_x)! (2m_y)!} \partial_x^{2 m_x} \partial_y^{2 m_y} \Delta^{(n)}_{xx}(t,x_0,y_0)(I_{n,m_x,m_y} - I_{n,m_x+1,m_y} ) \cr
   + & {1 \over (2m_x)! (2m_y)!} \partial_x^{2 m_x} \partial_y^{2 m_y} \Delta^{(n)}_{yy}(t,x_0,y_0) (I_{n,m_x,m_y} - I_{n,m_x,m_y+1} ) \cr
  - 2 & R^2 {1 \over (2m_x+1)! (2m_y+1)!}\partial_x^{2 m_x+1} \partial_y^{2 m_y+1} \Delta^{(n)}_{xy}(t,x_0,y_0)  I_{n,m_x+1,m_y+1} \Big\} \cr
\ea
\ee
The vanishing of the terms at order $R^{N+2}$ implies that
\beas
\Delta_{xx}^{(N)}(t,x_0,y_0) + \Delta_{yy}^{(N)}(t,x_0,y_0) &=& \sum_{(m_x,m_y) \ne (0,0)}  C^{N,m_x,m_y}_{xx} \partial_x^{2 m_x} \partial_y^{2 m_y} \Delta^{(N-2m_x -2m_y)}_{xx} \cr && \; \qquad \qquad + C^{N,m_x,m_y}_{yy}\partial_x^{2 m_x} \partial_y^{2 m_y} \Delta^{(N-2m_x -2m_y)}_{yy}  \cr && \; \qquad \qquad + C^{N,m_x,m_y}_{xy} \partial_x^{2 m_x - 1} \partial_y^{2 m_y - 1} \Delta^{(N-2m_x -2m_y)}_{xy}  \; ,
\eeas
where the $C$ coefficients can be read off from (\ref{ds2}). As examples, the first few equations give
\bea
\label{constraints}
\Delta_{xx}^{(0)}(t,x_0,y_0) + \Delta_{yy}^{(0)}(t,x_0,y_0) &=& 0 \cr
\Delta_{xx}^{(1)}(t,x_0,y_0) + \Delta_{yy}^{(1)}(t,x_0,y_0) &=& 0 \cr
\Delta_{xx}^{(2)}(t,x_0,y_0) + \Delta_{yy}^{(2)}(t,x_0,y_0) &=& -{1 \over 4} (\partial_y^2 \Delta_{xx}^{(0)}(t,x_0,y_0) + \partial_x^2 \Delta_{yy}^{(0)}(t,x_0,y_0)) \cr
&& - {3 \over 20} (\partial_x^2 \Delta_{xx}^{(0)}(t,x_0,y_0) + \partial_y^2 \Delta_{yy}^{(0)}(t,x_0,y_0)) \cr
&& + {1 \over 5} \partial_x \partial_y \Delta^{(0)}_{xy}(t,x_0,y_0) \cr
\Delta_{xx}^{(3)}(t,x_0,y_0) + \Delta_{yy}^{(3)}(t,x_0,y_0) &=& -{1 \over 6} (\partial_y^2 \Delta_{xx}^{(1)}(t,x_0,y_0) + \partial_x^2 \Delta_{yy}^{(1)}(t,x_0,y_0)) \cr
&& - {1 \over 6} (\partial_x^2 \Delta_{xx}^{(1)}(t,x_0,y_0) + \partial_y^2 \Delta_{yy}^{(1)}(t,x_0,y_0)) \cr
&& + {1 \over 9} \partial_x \partial_y \Delta^{(1)}_{xy}(t,x_0,y_0)
\eea
We see that this set of equations completely determines the combination $\Delta_{xx} + \Delta_{yy}$ at each order in $z$ in terms of the lower order terms in the expansion of $\Delta$. However, apart from the constraint (\ref{bound}) on the boundary behavior (equivalent to $\Delta_{\mu \nu}^{(0)} = 0$), the remaining elements of $\Delta_{\mu \nu}$ are completely unconstrained.

To constrain $\Delta_{\mu \nu}$ further, we need to use the requirement that the relation (\ref{dspert}) should hold in an arbitrary Lorentz frame. Thus, for each choice of reference frame, we will have equations analogous to (\ref{constraints}). Specifically, consider a general boost
\[
\Lambda = \left(\ba{ccc} \gamma & \gamma \beta_x & \gamma \beta_y \cr \gamma \beta_x & 1 + \beta_x^2 {\gamma^2 \over \gamma + 1} & \beta_x \beta_y {\gamma^2 \over \gamma + 1} \cr \gamma \beta_y & \beta_x \beta_y {\gamma^2 \over \gamma + 1} & 1 + \beta_y^2 {\gamma^2 \over \gamma + 1} \ea \right)
\]
In the equations for a general frame of reference obtained by such a boost, the left sides in (\ref{constraints}) will be replaced by
\[
\Lambda_x {}^\mu \Lambda_x {}^\nu \Delta_{\mu \nu} + \Lambda_y {}^\mu \Lambda_y {}^\nu \Delta_{\mu \nu} \; .
\]
Up to an overall constant factor, this gives
\[
\Delta_{ii} + 2 \beta_i \Delta_{it} + \beta^2 (\Delta_{tt} - {1 \over 2} \Delta_{ii}) + (\beta_i \beta_j - {1 \over 2} \delta_{ij} \beta^2)\Delta_{ij} \; .
\]
Now, consider the general version of the second equation in (\ref{constraints}) (the first equation already holds by (\ref{bound})). This requires the vanishing of
\[
\Delta^{(1)}_{ii} + 2 \beta_i \Delta^{(1)}_{it} + \beta^2 (\Delta^{(1)}_{tt} - {1 \over 2} \Delta^{(1)}_{ii}) + (\beta_i \beta_j - {1 \over 2} \delta_{ij} \beta^2) \Delta^{(1)}_{ij} \; .
\]
For a fixed $x_0$ and $y_0$, this is a polynomial in $\beta_i$ that must vanish for all values of $\beta_i$. Thus, the polynomial must be identically zero. At order $\beta^0$, this gives
\[
\Delta^{(1)}_{ii}(t,x_0,y_0) = 0
\]
as we had before. At order $\beta$, we get
\[
\Delta^{(1)}_{it}(t,x_0,y_0) = 0 \; .
\]
At order $\beta^2$, this gives
\[
\Delta^{(1)}_{tt}(t,x_0,y_0) = {1 \over 2} \Delta^{(1)}_{ii}(t,x_0,y_0) = 0
\]
and
\[
\Delta^{(1)}_{ij}(t,x_0,y_0) - {1 \over 2} \Delta_{ij} \Delta^{(1)}_{kk}(t,x_0,y_0) = 0 \; .
\]
Thus, we have $\Delta_{\mu \nu}^{(1)} = 0$. We can now continue to analyze the remaining equations in (\ref{constraints}) in turn. Supposing that we have shown $\Delta_{\mu \nu}^{(k)} = 0$ for $k < n$, the general version of the $n$th equation in (\ref{constraints}) requires the vanishing of
\[
\Delta^{(n)}_{ii} + 2 \beta_i \Delta^{(n)}_{it} + \beta^2 (\Delta^{(n)}_{tt} - {1 \over 2} \Delta^{(n)}_{ii}) + (\beta_i \beta_j - {1 \over 2} \delta_{ij} \beta^2) \Delta^{(n)}_{ij} \; ,
\]
since the right hand side in (\ref{constraints}) will be zero. Repeating the analysis above, we conclude that $\Delta_{\mu \nu}^{(n)} = 0$. By induction, this holds for all $n$, so we have shown that $\Delta_{\mu \nu} = 0$, completing the proof.

\section{Discussion}

In this paper, we have seen that to linear order in perturbations about the vacuum state, the emergence of gravitational dynamics in the theory dual to a holographic CFT is directly related to a general relation satisfied by CFT entanglement entropies on ball-shaped regions. This relation is closely related to the First Law of Thermodynamics, but is more general since it applies to arbitrary perturbations of the state rather than perturbations for which the system remains in thermal equilibrium.

Some of the derivations we have provided are specific to the case of four-dimensional gravity. However, the proof given in \cite{Blanco:2013joa} that Einstein's equations imply $\delta S = \delta E$, and our proofs in section 5.3 and appendix A that $\delta S = \delta E$ implies the linearized Einstein's equations, are valid in any number of dimensions.

The linearized Einstein's equations we derived are for the metric components in the field theory directions and radial direction of the bulk. Any additional fields in the gravitational theory, including metric components in any compactified directions, are not constrained by the CFT relation we have considered. At linear order, the equations for these fields decouple from the linearized Einstein's equations for the metric in the non-compact directions. Thus, we can say that the universal relation $\delta S = \delta E$ is equivalent to the universal sector of the linearized bulk equations.

Our results do not imply that all holographic theories are dual to gravitational theories whose metric perturbations satisfy Einstein's equations. In this paper, we assumed that entanglement entropies are related to areas via the usual  Ryu-Takayanagi formula, and that the stress-energy tensor in the dual field theory is related to the asymptotic form of the metric. In more general theories, the entanglement entropy may correspond to a more complicated functional of the bulk geometry and the relation between the stress tensor and asymptotic metric may be modified. In these cases, we expect that the bulk equations will be different, for example involving $\alpha'$ corrections with higher-derivative terms. However, it may be possible following the methods in this paper to derive the linearized version of these more general equations given a particular choice for the holographic entanglement entropy formula and the holographic formula for the stress tensor.

It will be interesting to see whether the first non-linear corrections to Einstein's equations in the bulk are equivalent to some simple property of entanglement entropies.

\section*{Acknowledgments}

This research is supported in part by the Natural Sciences and Engineering Research Council of Canada.

\appendix

\section{Alternative derivation of linearized Einstein's equations from $\delta E= \delta S$}

In this appendix, we offer an alternative proof that solutions of Einstein's equations satisfy $\delta S_A = \delta E^{hyp}_A$. This proof replaces $\delta S_A = \delta E^{hyp}_A$ with the infinite set of relations obtained by matching the terms in the power series expansion of this relation in $R$, the radius of the disk $A$, as we did in section 4.2.

\subsection{Expansion of $\delta E = \delta S$ in powers of $R$}

To begin, we expand both (\ref{LS}) and (\ref{RS}) in powers of $R$. Defining
\be
\label{Hexp}
H_{\mu \nu}(z,x,y) = \sum_{n=0}^\infty z^n H_{\mu \nu}^{(n)}(x,y)
\ee
we have
\be
\label{RS2}
\delta E = {3 \over 2} \sum_{m_x,m_y=0} R^{2 + 2 m_x + 2 m_y} I_{2,m_x,m_y} \partial_x^{2 m_x} \partial_y^{2 m_y} H^{(0)}_{tt}(t,x_0,y_0) \;
\ee
while
\be
\label{LS2}
\ba{rl} \delta S =  \sum R^{n + 2 m_x + 2 m_y + 2} \Big\{ &    {1  \over (2m_x)! (2m_y)!} \partial_x^{2 m_x} \partial_y^{2 m_y} H^{(n)}_{xx}(t,x_0,y_0)(I_{n,m_x,m_y} - I_{n,m_x+1,m_y} ) \cr
   + & {1 \over (2m_x)! (2m_y)!} \partial_x^{2 m_x} \partial_y^{2 m_y} H^{(n)}_{yy}(t,x_0,y_0) (I_{n,m_x,m_y} - I_{n,m_x,m_y+1} ) \cr
  - 2 & R^2 {1 \over (2m_x+1)! (2m_y+1)!}\partial_x^{2 m_x+1} \partial_y^{2 m_y+1} H^{(n)}_{xy}(t,x_0,y_0)  I_{n,m_x+1,m_y+1} \Big\} \cr
\ea
\ee
where $I$ was defined in (\ref{defI}).

\subsection{Checking that solutions of Einstein's equations satisfy $\delta S = \delta E$}

Using these expansions, it is straightforward to verify that any solution of the linearized Einstein's equations (\ref{Einstein}) satisfies $\delta E = \delta S$, as was done originally in \cite{Blanco:2013joa} and by another alternative approach in section 4.

Using the expansion (\ref{Hexp}), the equations (\ref{Einstein}) become
\bea
\label{Einstein2}
H_{tt}^{(n)} &=& H_{xx}^{(n)} + H_{yy}^{(n)}  \label{EE1}\\
\partial_t H^{(n)}_{tt} &=& \partial_x H^{(n)}_{tx} + \partial_y H^{(n)}_{ty} \label{EE2}\\
\partial_t H^{(n)}_{tx} &=& \partial_x H^{(n)}_{xx} + \partial_y H^{(n)}_{xy} \label{EE3}\\
\partial_t H^{(n)}_{ty} &=& \partial_x H^{(n)}_{xy} + \partial_y H^{(n)}_{yy} \label{EE4}\\
 H^{(n)}_{\mu \nu} &=& {1 \over n (n+3)} (\partial_t^2 - \partial_x^2 - \partial_y^2) H_{\mu \nu}^{(n-2)} \qquad n \ge 2 \label{EE5} \\
 H^{(1)}_{\mu \nu} &=& 0 \label{EE6} \; .
\eea
Starting with (\ref{EE1}) and then using (\ref{EE2}), (\ref{EE3}), (\ref{EE4}), and finally (\ref{EE5}), we find:
\[
\ba{rl}
& \partial_t^2 H_{tt}^{(n)} = \partial_t^2 (H_{xx}^{(n)} + H_{yy}^{(n)}) \qquad \qquad  \cr
\Rightarrow & \partial_t (\partial_x H_{xt}^{(n)} + \partial_y H_{yt}^{(n)})  = \partial_t^2 (H_{xx}^{(n)} + H_{yy}^{(n)})  \cr
\Rightarrow & \partial_x^2 H_{xx}^{(n)} + \partial_y^2 H_{yy}^{(n)} + 2 \partial_x \partial_y H_{xy}^{(n)}  = \partial_t^2 (H_{xx}^{(n)} + H_{yy}^{(n)})   \cr
\Rightarrow & \partial_x^2 H_{xx}^{(n)} + \partial_y^2 H_{yy}^{(n)} + 2 \partial_x \partial_y H_{xy}^{(n)}  = \partial_t^2 (H_{xx}^{(n)} + H_{yy}^{(n)})   \cr
\Rightarrow & \partial_x^2 H_{xx}^{(n)} + \partial_y^2 H_{yy}^{(n)} + 2 \partial_x \partial_y H_{xy}^{(n)}  = (\partial_x^2 + \partial_y^2) (H_{xx}^{(n)} + H_{yy}^{(n)}) + (n+2)(n+5) (H_{xx}^{(n+2)} + H_{yy}^{(n+2)})   \cr
\Rightarrow & 2 \partial_x \partial_y H_{xy}^{(n)}  = \partial_y^2 H_{xx}^{(n)} + \partial_x^2 H_{yy}^{(n)} + (n+2)(n+5) (H_{xx}^{(n+2)} + H_{yy}^{(n+2)})   \cr
\ea
\]
Using this last equation, we can eliminate $H_{xy}^{(n)}$ from (\ref{LS2}). This gives
\bea
\label{LS3}
\delta S &=&  \sum R^{n + 2 m_x + 2 m_y + 2} \Big\{  {1  \over (2m_x)! (2m_y)!} \partial_x^{2 m_x} \partial_y^{2 m_y} H^{(n)}_{xx}(t,x_0,y_0) C^{xx}_{n,m_x,m_y} \cr
   && +  {1 \over (2m_x)! (2m_y)!} \partial_x^{2 m_x} \partial_y^{2 m_y} H^{(n)}_{yy}(t,x_0,y_0) C^{yy}_{n,m_x,m_y} \Big\}
\eea
where for $n \ge 2$ we have
\beas
C^{xx}_{n,m_x,m_y} &=& I_{n,m_x,m_y} - I_{n,m_x+1,m_y} - {2 m_y \over 2 m_x + 1} I_{n,m_x+1,m_y} - {n(n+3) \over (2m_x+1)(2m_y+1)} I_{n-2,m_x+1,m_y+1} \cr
&=& 0 \cr
C^{yy}_{n,m_x,m_y} &=& I_{n,m_x,m_y} - I_{n,m_x,m_y+1} - {2 m_x \over 2 m_y + 1} I_{n,m_x,m_y+1} - {n(n+3) \over (2m_x+1)(2m_y+1)} I_{n-2,m_x+1,m_y+1} \cr
&=& 0
\eeas
while for $n=1$ and $n=0$, we have
\beas
C^{xx}_{1,m_x,m_y} &=& I_{1,m_x,m_y} - I_{1,m_x+1,m_y} - {2 m_y \over 2 m_x + 1} I_{1,m_x+1,m_y}
= {4 \over 3} I_{3,m_x,m_y} \cr
C^{yy}_{1,m_x,m_y} &=& I_{1,m_x,m_y} - I_{1,m_x,m_y+1} - {2 m_x \over 2 m_y + 1} I_{1,m_x,m_y+1}
= {4 \over 3} I_{3,m_x,m_y}
\eeas
and
\beas
C^{xx}_{0,m_x,m_y} &=& I_{0,m_x,m_y} - I_{0,m_x+1,m_y} - {2 m_y \over 2 m_x + 1} I_{0,m_x+1,m_y}
= {3 \over 2} I_{2,m_x,m_y} \cr
C^{yy}_{0,m_x,m_y} &=& I_{0,m_x,m_y} - I_{0,m_x,m_y+1} - {2 m_x \over 2 m_y + 1} I_{0,m_x,m_y+1}
= {3 \over 2} I_{2,m_x,m_y} \; .
\eeas
In each case, we have made simplifications using the definition (\ref{defI}) of $I$. Using these results together with (\ref{EE6}), we find that (\ref{LS3}) simplifies to
\beas
\label{LS3}
\delta S &=&  \sum R^{2 m_x + 2 m_y + 2}   {1  \over (2m_x)! (2m_y)!} \partial_x^{2 m_x} \partial_y^{2 m_y} (H^{(0)}_{xx}(t,x_0,y_0) + H^{(0)}_{yy}(t,x_0,y_0)) ({3 \over 2} I_{2,m_x,m_y}) \cr
&=& {3 \over 2} \sum R^{2 m_x + 2 m_y + 2}   {1  \over (2m_x)! (2m_y)!} \partial_x^{2 m_x} \partial_y^{2 m_y} H^{(0)}_{tt}(t,x_0,y_0)  I_{2,m_x,m_y} \cr
&=& \delta E
\eeas
Thus, we have verified that $\delta S = \delta E$ for linearized solutions of Einstein's equations, providing an alternate argument to the one in \cite{Blanco:2013joa}.

\end{document}